\documentclass[11pt]{article}%
\usepackage{color}
\usepackage{times,graphics,float,subfig}
\usepackage{enumerate}
\usepackage{amsfonts,amsthm,amscd}
\usepackage{amssymb}
\usepackage{graphicx}
\usepackage{rotating}
\usepackage{acronym}
\usepackage{amsmath}
\usepackage{helvet}
\usepackage{courier}
\usepackage{type1cm}
\usepackage{setspace}
\usepackage[subfigure]{tocloft}

\usepackage{makeidx}         
\usepackage{multicol}        
\usepackage[bottom]{footmisc}
\usepackage{epsfig}
\usepackage{mathrsfs}
\usepackage{index}
\usepackage[all]{xy}
\usepackage{morefloats}
\usepackage{emptypage}
\newtheorem{theorem}{Theorem}

\newtheorem{corollary}[theorem]{Corollary}

\newtheorem{definition}[theorem]{Definition}

\newtheorem{remark}[theorem]{Remark}

\usepackage{hyperref}

\begin{document}
\title{A Phylogenetic Trees Analysis of SARS-CoV-2}
\author{Chen Shen, Vic Patrangenaru and Roland Moore}
\date{June 13, 2021}
\maketitle

\begin{abstract} One regards spaces of trees as stratified spaces, to study distributions of phylogenetic trees. Stratified spaces with may have cycles, however spaces of trees with a fixed number of leafs are contractible. Spaces of trees with three leafs, in particular, are spiders with three legs. One gives an elementary proof of the stickiness of intrinsic sample means on spiders. One also represents four leafs tree data in terms of an associated Petersen graph. One applies such ideas to analyze RNA sequences of SARS-CoV-2 from multiple sources, by building samples of trees and running nonparametric statistics for intrinsic means on tree spaces with three and four leafs. SARS-CoV-2 are also used to built trees with leaves consisting in addition to other related coronaviruses.
\end{abstract}


\section{Introduction}

We first introduce some basic notions of phylogenetics that are common knowledge in Biology. A phylogenetic tree is a directed tree ({directed simply connected graph}), having a distinguished vertex, that is an ancestor of all other vertices, called {\bf root}; all vertices that have no descendants are called {\bf leaves} of the phylogenetic tree. In a phylogenetic tree, each edge is called a {\bf branch.} Each vertex where the two branches meet is called a {\bf branching point}, or node. A node is the most recent common ancestor of all species on the other vertices of those branches.

Deoxyribonucleic acid (DNA) sequences can be used to draw a phylogenetic tree. A nucleotide consists of a sugar molecule (either ribose in ribonucleic acid (RNA) or DNA attached to a phosphate group and a nitrogen-containing base. The bases used in DNA are purines adenine (A) and guanine (G), pyrimidine cytosine (C) and pyrimidine thymine (T). In RNA, the base pyrimidine uracil (U) takes the place of T. To construct a tree, we will compare the RNA or DNA sequences of different species or individuals. Related individuals have a common ancestor. Before species split into separate descendants, they had the same RNA (or DNA). But as species evolve and diverge, they will accumulate changes in the DNA sequences. We can use these changes in the DNA to tell how closely related two individuals are. If there are not very many differences, they are probably closely related; if there are many changes, they might be distant relatives. \\
The DNA sequences are double-stranded {and are} made of letters A, G, C, and T, and RNA sequences {are similar to those of the DNA and} made of letters A, G, C, and U. Chromosomes are thread-like structures located inside the nucleus of animal and plant cells. Each chromosome is made of protein and a single molecule of DNA. Passed from parents to offspring, DNA contains the specific instructions that make each type of living creature unique. A gene is an ordered sequence of nucleotides located in a particular position on a particular chromosome that encodes a specific functional product (i.e., a protein or RNA molecule). Biologists commonly use one {homogeneous} sequence, which in the biologist's language concerns the relationship between gene trees and genes that are made from one tree. The gene sequence might be about 200 base pairs long. One of the problems that has occurred in the last fifty years is that biologists believe that the way evolution works is that there would only be one {\em species tree}. Different genes have different histories, so you get different gene trees. Putting them together is a statistical problem that helps study of the evolutionary process.

A phylogenetic tree with $p$ leaves is an equivalence class based on a certain equivalence of a DNA-based connected directed graph of species with no loops, having an unobserved {\em root} (common ancestor) and $p$ observed {\em leaves} (currently observed species of a certain family of living creatures).
The paper is organized as follows; in the first section one introduces tree spaces of trees with $m$ leafs, which are regarded as stratified spaces of dimension $m-2$. In particular the detailed description is fiven for the spaces $T_3$ and $T_4$. It is shown that $T_3$ is a 3-spider, and $T_4$ can be described in terms of the so called Petersen graph.
Section three is dedicated to a proof of the central limit theorem for intrinsic means of spiders. An elementary proof of the stickiness theorem (see Hotz et al (2013)\cite{HoHuLeMaMaMiNoOwPaSk:2013} is given here.  The fourth section is dedicated to the CLT for intrinsic means on $T_4$ (see Bardem et al.(2013)\cite{BaLeOw:2013}). Here the key result is the partial stickiness of the intrinsic means, that stick to the 1D stratum of the tree space.
The last part of the paper is applied, classifying SARS-CoV-2 data, according to stickiness of their intrinsic sample means.

\section{Tree Spaces}

Our stratified data analysis requires certain basic concepts, that are fairly recent in Statistics (see eg Patrangenaru and Ellingson (2015)\cite{PaEl:2015}, p.475, and references there in).
\begin{definition} A stratified space (space with a manifold stratification) is a metric space $\mathcal{M}$ that admits a filtration $\emptyset= F_{-1} \subseteq F_0 \subseteq F_1 \dots \subseteq F_n \subset \dots =M = \cup_i F_i$, By closed subspaces, such that for each $ i=1,\dots,n, F_i\setminus F_{i-1}$ is empty or is an $i$-dimensional manifold, called the $i$-th stratum.
\end{definition}
An elementary example of a 2 dimensional stratified space is a cone $\mathcal C,$ set of solutions in $\mathbb R^3$ of the equation $x^2+y^2-z^2=0;$ In this case $F_0=\{(0,0,0)\}, F_2= \mathcal C, F_i\setminus F_{i-1}=\empty,\forall i \notin {0,2}.$

A tree with $p$ leaves is a connected, simply connected graph , with
a distinguished vertex, labeled $o$, called the root, and $p$ vertices of
degree 1, called leaves, that are labeled from 1 to $p$. In addition, we
assume that with all interior edges have positive lengths. (An edge
of a p-tree is called interior if it is not connected to a leaf.)

Now consider a tree $T$, with interior edges $e_1,...,e_r$ of lengths
$l_1,...,l_r$, respectively. If $T$ is binary, then $r = p - 2$, otherwise $r < p - 2$. The vector${(l_1,...,l_r)}^T$ specifies a point in the positive open orthant ${(0,\infty)}^r$.

An $p$-tree has the maximal possible number of interior edges $p-2$ and thus determines the largest possible dimensional orthant, when it is a binary tree; in this case the orthant is $p-2$ dimensional. The open orthants form the highest dimensional stratum $ F_{p-2}\setminus F_{p-3}$ of $T_p,$ regarded as a $p-2$ dimensional stratified space.  The orthant corresponding to each tree which is not binary appears as a boundary face of the orthants corresponding to at least three binary trees; in particular the origin of each orthant corresponds to the (unique) tree with no interior edges. We construct the space $T_p$ by taking one $p-2$ dimensional orthant for each of the $(2p-3)!!$ possible binary trees, and gluing them together along their common faces. Due to such gluing operations, tree spaces are {\it not} manifolds. Singularities (points where the space does not have a tangent space of the same dimension as the space itself) are present in the tree space structure. For further details on the construction of the tree space $T_p$, see Billera {\em et al.}(2001)\cite{BiHoVo:2001};

\begin{figure}[H]
\centering
\label{f:tree-spaces}
\includegraphics[scale = 0.27]{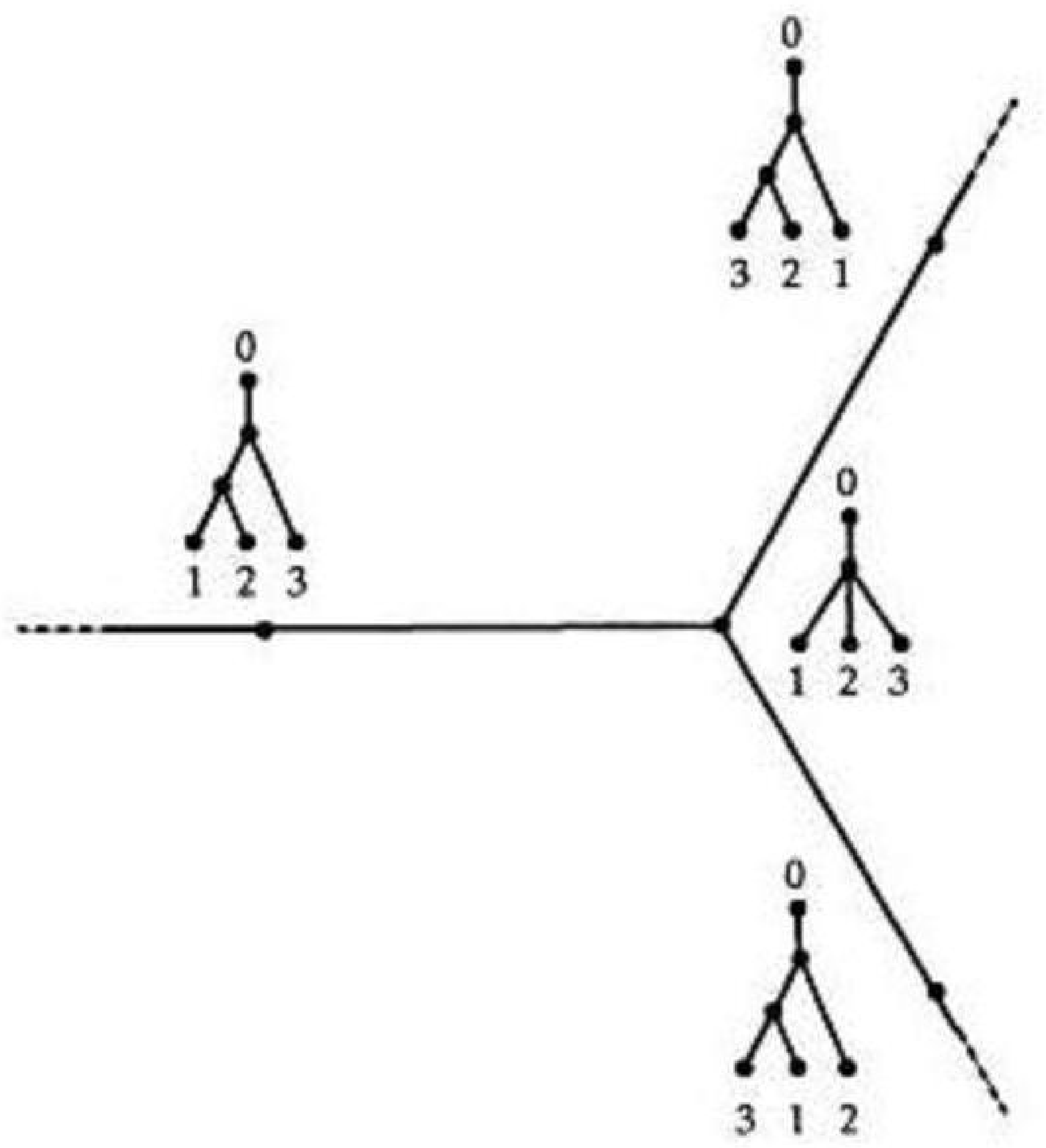}
\includegraphics[scale = 0.27]{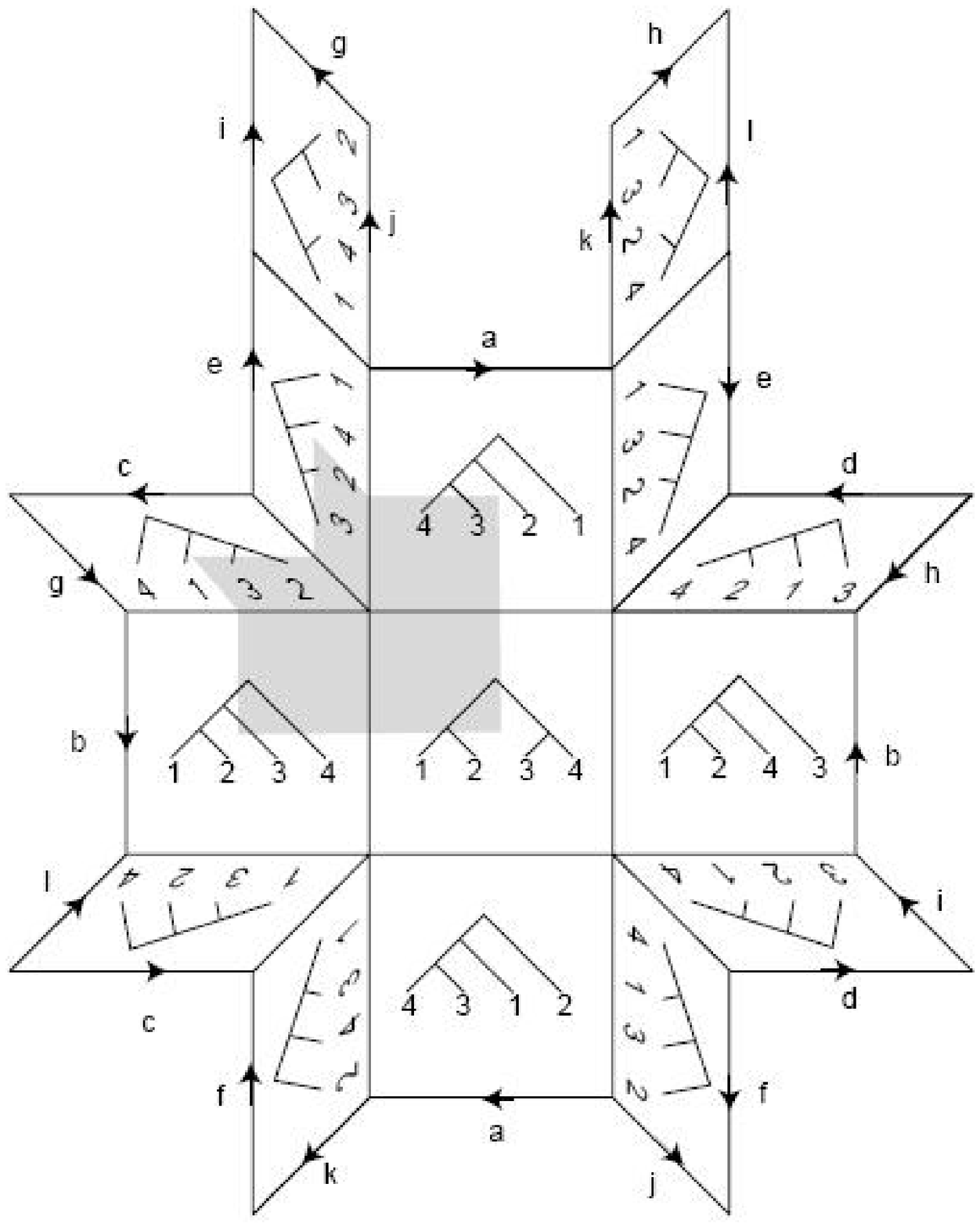}
\caption{Tree spaces $T_3$ and $T_4$.} \label{f:ts}
\end{figure}

In this paper, we are considering phylogenetic trees with 3 or 4 leafs. The space of trees with 3 leafs is $T_3$ can be identified with $S_3$, a $3$-spider,  which is the union of three line segments with a common end (see left hand side of Figure \ref{f:ts}).

$T_4$ as a stratified space has $15 = (2\times 4 - 3)!!$ 2D quadrants glued according to tree identification rules (see Billera et. al.(2001)\cite{BiHoVo:2001}). Interior points of these quadrants are combinatorial binary trees with four leaves, the coordinates of an interior point being given by the two interior edges of a binary tree in one of these combinatorial binary trees. Points on the boundaries of the quadrants, are on the one dimensional stratum, and correspond to combinatorial trees with four leaves, which are obtained from a  combinatorial binary tree by shrinking one of the interior edges to zero length (see right hand side of Figure \ref{f:ts}). The zero dimensional stratum is made of one point only, the {\em star tree}.

\begin{figure}[H]
\begin{center}
\vspace{-0.45cm}
\includegraphics[scale=0.6]{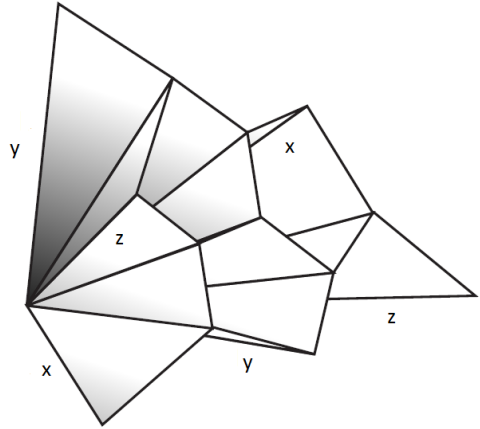}
\caption{A 2D stratified space - $T_4$, space of trees with four leaves} \label{f:t4}
\end{center}
\end{figure}

\begin{figure}[H]
\begin{center}
\vspace{-0.45cm}
\includegraphics[scale=0.4]{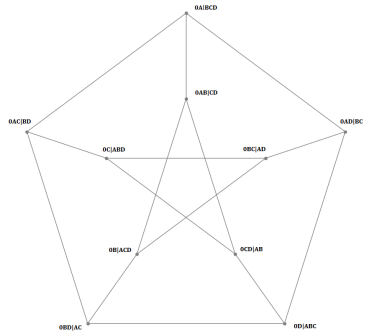}
\caption{Petersen graph} \label{f:Petersen}
\end{center}
\end{figure}

 Therefore a representation of $T_4$ as a surface with singularities can be obtained from the polyhedral surface obtained by the identifications of three semi-axes labeled with the same letter as shown in Figure \ref{f:t4}. While this is a 3D pictorial representation only, in general $T_m$ can be embedded in $\mathbb R\textcolor{blue}{^{k}}$, $k=2^m-m-2$ ( see Ellingson et al(2014) \cite{ElHePaVa:2014}), so, in fact, given that the number of edges that form the one dimensional stratum of $T_4$ is $2^4 - 4 - 2 = 10,$ this stratified space is embedded in $\mathbb R^{10}$ having the {star tree} at the origin. In this representation, the intersection of a sphere in $\mathbb R^{10}$ centered at the origin with $T_4$ is the so called {\em Petersen graph}. The embedded surface can be thus regarded as a one sheet cone over the Petersen graph. An edge of the Petersen graph is the transverse intersection of one quadrant with the sphere; thus there are $15$ edges; a vertex of this graph is a point where one of the coordinate semi-axes pierces the sphere; therefore there are $10$ vertices ( see figure \ref{f:Petersen}).

\section{An elementary proof of the intrinsic CLT on spiders}

The intrinsic mean on $T_m,$ with the piecewise Euclidean metric is unique, since $T_m$ with this metric is a CAT(0) space. A CLT for the intrinsic sample mean on a tree space $T_m$, for $m$ fixed, is however still ongoing research. The most recent results, to our knowledge, are due to Shen (2021)\cite{Sh:2021}, Barden et al.(2018)\cite{BaLeOw:2018}, and Barden and Le(2018)\cite{BaLe:2018} are are concerned only with the case when the intrinsic mean lies on the first and second highest dimensional strata of $T_m.$
In this section we are concerned with the case $m=3,$ when it is known that $T_3=S_3.$ Therefore we study at no additional effort, that case of the CLT for the intrinsic sample mean on a spider $S_p.$
Assume $X_i, i =1, \dots, n$ are i.i.d. random objects on a spider $S_p,$ having legs $L_a, a = 1, \dots p,$ and center $C.$ Further, assume the intrinsic mean $\mu_{X_1,I}$ exists and the intrinsic variance is finite. Any probability measure $Q$ on $S_p$ decomposes uniquely as a weighted sum of probability measures $Q_k$ on the legs $L_k$ and an atom $Q_0$ at $C$ (Hotz et. al.(2013)\cite{HoHuLeMaMaMiNoOwPaSk:2013}).  More precisely, there are nonnegative real numbers $\{w_k\}_{k=0}^p$ summing to~$1$ such that, for any Borel set $A
\subseteq S_p$, the measure $Q$ takes the value
\begin{equation}\label{e:prob-meas-on-spider}
  Q(A) = w_0 Q_0(A \cap C) + \sum_{k=1}^p w_k Q_k(A \cap L_k).
\end{equation}
We will consider the nontrivial case when the moments $ \nu_a = E(Q_a), a = 1, \dots, p$ are all positive.
\begin{theorem}\label{t:clt-on-spider}(Hotz et. al. (2013)). Assume $w_0 = 0.$ (i) If there exists $a \in \overline{1,p}$ such that $w_a\nu_a > \sum_{b \neq a}w_b \nu_b$ then $\mu_{X_1,I} \in L_a$ and, for $n$ large enough $\bar X_{n,I}\in L_a $ and $\sqrt{n}(\bar X_{n,I}-\mu_{X_1,I})$ has asymptotically a normal distribution.
(ii) If there exists $ a \in \overline{1,p}$ such that $w_a\nu_a = \sum_{b \neq a}w_b \nu_b,$ then, after folding the legs $L_b, b\neq a,$ into one half line opposite to $L_a$,
$\sqrt{n}(\bar X_{n,I})$ has asymptotically the distribution of the absolute value of a normal distribution.
(iii) If $\forall a \in \overline{1,p}, w_a\nu_a < \sum_{b \neq a}w_b \nu_b,$ then $\mu_{X_1,I}=C$ and there is $n_0$ s.t. $\forall n \ge n_0,$ then $ \bar X_{n,I}$ = $C$ a.s.
\end{theorem}
For a {\bf proof}, recall that if $x\in L_a$, the Fr\'echet function $F$ is defined as follows,
\begin{align}\label{e:frechet-spider}
 F(x) &= \sum_{i=1,i\neq a}^K \int_0^{\infty} (x+u)^2w_iQ_i(du) + \int_0^{\infty}(x-u)^2w_aQ_a(du) \\
 &= x^2\sum_{i=1}^K \int_0^{\infty}w_iQ_i(du) +2x[\sum_{i=1,i\neq a}^K  \int_0^{\infty} uw_iQ_i(du) -  \int_0^{\infty} uw_aQ_a(du)]+ \sum_{i=1}^K  \int_0^{\infty}u^2w_iQ_i(du)\\
 &=x^2+2[\sum_{i=1,i\neq a}^K v_i - v_a]x +const.
\end{align}
where $v_i= \int_0^{\infty}uw_iQ_i(du)=w_i\nu_i.$

Since there exists an unique minimizer for the Fr\'echet function $F$ in \eqref{e:frechet-spider}, this minimizer is intrinsic mean $\mu_I$. The minimizer of this quadratic function is $x^*=v_a-\sum_{i\neq a}v_i$, where $x \in L_a =\{(a,u):u\in [0,\infty)\}$. Thus, we have three situations:(i)$v_a-\sum_{i\neq a}v_i>0$ or $v_a>\sum_{i\neq a}v_i$, (ii)$v_a-\sum_{i\neq a}v_i=0$ or $v_a=\sum_{i\neq a}v_i$ and {iii} $v_a-\sum_{i\neq a}v_i<0$ or $v_a<\sum_{i\neq a}v_i$. In case(i), we have $\mu_I$ well defined on $L_a$, then classical C.L.T is applied. In case (ii), we can fold other legs into that half line opposite to $L_a$ then apply C.L.T. and since the negative part is undefined, so the result goes to a positive truncated normal distribution. In case (iii), for any $a\in \{1,\dots,K\}$, we have $\mu_I=C$, which shows that intrinsic mean $\mu_I$ sticks to the center $C$;

\begin{corollary} Under the assumptions of Theorem \ref{t:clt-on-spider} (iii), we say that the sample mean is {\em sticky}, a condition that can be readily verified for data on a spider.
\end{corollary}

\begin{remark} The first paper on the asymptotic behavior of the CLT on a stratified space, is due to Basrak(2010)\cite{Ba:2010}, who gave the
first stickiness example for intrinsic means of distributions on metric binary trees. According to Basrak(2010):``
 Limit theorems on ( binary ) trees will need minor adjustments, since on a general tree, the barycenter can split the tree into more than three subtrees. Nevertheless, asymptotically, the inductive mean will have one of the three types of behavior described in Theorem 3 in Basrak(2010), meaning that the stickiness phenomenon is still present for distributions on trees. This comment from Basrak(2010) does not extend to arbitrary finite, connected graphs, since graphs usually have cycles (see Figure \ref{f:graph}).

\begin{figure}[b]
\begin{center}
\includegraphics[scale = 0.6]{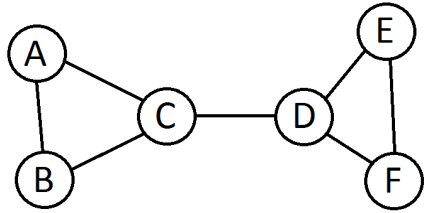}
\caption{Bicyclic graph } \label{f:graph}
\end{center}
\end{figure}
\end{remark}

However, Basrak's condition for trees can be extended to the case of graphs in {\em some} general cases even if the graph $G$ has cycles with positive mass on any arc of a cycle. Given a graph $G,$ for each $x\in G$ and leg $L_a$, $d(x,\xi)$ is either increasing or decreasing for $\xi\in L_a$.
We define $\varepsilon_a(x)=+1$ if increasing and $\varepsilon_a(x)=-1$ if decreasing.
Then $\mu_I$ is sticky if for all legs $L_a$
\begin{equation}\label{sticky} E(d(X,\mu_I)\varepsilon_a(X))>0.
\end{equation}
This is easily identified as a condition that implies that $\mu_I$ is a local minimum
of the Fr\'echet function. The quantity $m_a= E(d(X,\mu_I)\varepsilon_a(X))$ is called
the {\em net moment in the direction of } $L_a$.
\\
As an aside, one can consider for any $\xi\in G$ its star neigborhood $S_\xi$ and the ``derivative''
in the direction of leg $L_a$ of $S_\xi,$ given by
\begin{equation}\label{derivative}D_{L_a}F_d = E(2d(X,\xi)\varepsilon_a(X)).
\end{equation}
Returning to the Fr\'echet mean $\mu_I$, it follows that for large $n$ and samples $X_1,\ldots,X_n$ and leg $L_a,$ then
\begin{equation}\label{e:limit-h}\frac1n\sum_{i=1}^n d(X_i,\mu_I)\varepsilon_a(X_i)\to_d E(d(X,\mu_I)\varepsilon_a(X))>0 \hbox{ almost surely.}
\end{equation}
Therefore, as in Theorem\ref{t:clt-on-spider}(iii), with high probability $\mu_I$ is a local minimum for the Fr\'echet function corresponding to the sample.
One should show that with high probability $\mu_I$ is a point of absolute minimum for the Fr\'echet function corresponding to the empirical.
\begin{remark} (Hendriks(2014)) One may note that in the case of graphs having cycles, the condition of stickiness is similar with the condition in Theorem \ref{t:clt-on-spider}. Consider the situation where $\mu_I$ is a vertex, and $C_a$ is a connected component of $G\backslash\{\mu_I\}$ that has a unique edge joined to $\mu_I.$
Then $G_a=C_a\cup\mu_I$ is a subgraph of $G$. For each $x\in G_a$ there is the intrinsic distance from
$x$ to $\mu_I$, and with respect to the probability $Q_a$ conditional to be in $C_a$ there is an
expected intrinsic distance, called the moment (of $Q_a$ or $G_a$ with respect to $\mu_I$), which we label $m_a.$ The condition for stickiness is equivalent in this case to
$m_a < \sum_{b \neq a}m_a,$ which in the case of a star tree (spider), neighborhood of the Fr\'echet mean, is the right notion to study the empirical limit behavior, equivalent to the condition in Theorem \ref{t:clt-on-spider} (iii).
\end{remark}
\begin{remark} Certain complications arise in describing stickiness phenomena, in case
when the graph includes a cycle of positive probability mass. Consider for example the case
of a simple graph such as the circle, realizable as a one vertex, one edge graph.
In this case, if $\mu_I$ is a Fr\'echet mean, then the antipodal point, $-\mu_I$, must have probability $0.$
Even, if the density in a neighborhood of $-\mu_I$ is continuous, then the density at $-\mu_I$
cannot exceed $(2\pi)^{-1}$.
If the density is below $(2\pi)^{-1}$ Hotz and Huckemann (2015)\cite{HoHu:2015} prove a central limit theorem to a distribution other than normal.
\end{remark}

\section{On the CLT on $T_4$}

\begin{figure}[H]
\begin{center}
\includegraphics[scale = 0.6]{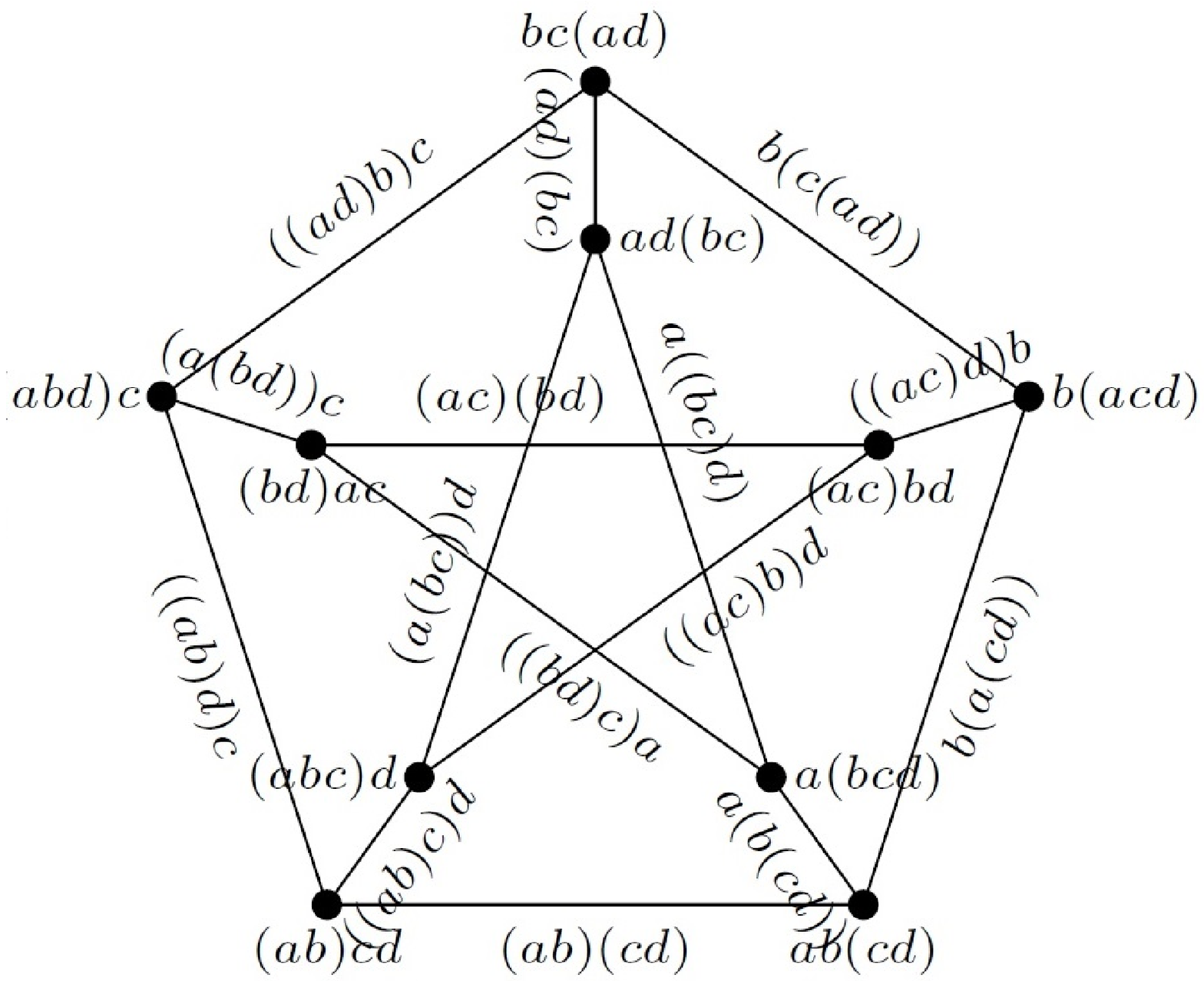}
\end{center}
\caption{Structure of $T_4$ reflected on the Petersen graph.}
\label{f:t4a}%
\end{figure}

The space $T_4$ consists of fifteen 2-dimensional faces that are quadrants of planes bounded by certain pairs of the out of the ten positive semi-axes in $\mathbb R^10,$ that are dictated by the leaf tree structure, as shown on the Petersen graph in Figure \ref{f:t4a} (see Barden et al(2013)\cite{BaLeOw:2013}).

\begin{figure}[H]
\centering
\begin{center}
\includegraphics[scale = 0.6]{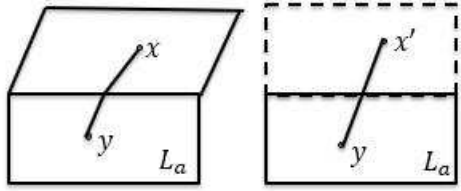}
\end{center}
\caption{Distance between points on different leafs of an open book}%
\label{fig:epsilon-disk}%
\end{figure}

There are a number of cases to consider, when it comes to CLT on $T_4$, depending on whether the population intrinsic mean lies in the 2D stratum, in the 1D stratum or at the origin (star tree). The limiting distribution for all cases are given in Barden et al (2013)\cite{BaLeOw:2013}. In particular here we detail the case, when the data lies on three quadrants forming what is known as an {\em open book} with three leafs (see Hotz el al(2013)\cite{HoHuLeMaMaMiNoOwPaSk:2013}), obtained by gluing these quadrants along a common boundary (spine). Such a particular type of open book could be defined as $L_i=\{(i,x_1,x_2):x_1,x_2\in [0,\infty)\} i=1,2,3$, together at the common spine $S=[0,\infty)$ which comprises the equivalence classes in $\cup_{i=1}^3([0,\infty)\times [0,\infty) \times \{i\}$. Thus, the open book $O_3$ is the disjoint union
\[
O_3 = S \cup L_1^+ \cup L_2^+ \cup L_3^+
\]
of the spine $S$ and the interiors $L_i^+ = L_i \slash S$ of the leaves, $i=1,2,3$. If the points $\textbf{x}=(x_1,x_2),\textbf{y}=(y_1,y_2)$ are on the same flat convex leaf (as a subset of $\mathbb R^2,$  the distance between them $d(\textbf{x},\textbf{y})=\|\textbf{x}-\textbf{y}\|$. If the points $\textbf{x},\textbf{y}$ are on different leafs (see figure \ref{fig:epsilon-disk}), one could replace point $\textbf{x}$ by $\textbf{x'}$ onto the half space opposite to $L_a$, then the distance is defined as $d(\textbf{x},\textbf{y})=\|\textbf{x'}-\textbf{y}\|$, where $\textbf{x'}=(x_1,-x_2)$.

Assume $X_i, i=1,\dots,n$ are i.i.d. random objects on $O_3$. Denote a weighted sum of probability measure $Q_3$ on the legs $L_i$ and an atom $Q_0$ at $C$. Further, note that the intrinsic mean $\mu_I$ exists since space has  and the intrinsic variance is finite. We assume $w_0=0$ and $x\in L_a$, the Fr\'echet function is defined as follows,

\begin{align*}
  F(\textbf{x}) & =  \sum_{i=1,i\neq a}^3 \int_{O_3} \|\textbf{x'}-\textbf{u}\|^2w_iQ_i(d\textbf{u})+ \int_{O_3} \|\textbf{x}-\textbf{u}\|^2w_aQ_a(d\textbf{u}) \\
   & =\|\textbf{x}\|^2-\sum_{i\neq a} 2\int_o^{\infty}\langle \textbf{x'},\textbf{u}\rangle w_iQ_i(d\textbf{u})-2\int_o^{\infty}\langle \textbf{x},\textbf{u}\rangle w_aQ_a(d\textbf{u})+\sum_{i=1}^K \int_0^{\infty}\|\textbf{u}\|^2w_iQ_i(d\textbf{u})\\
   & =x_1^2-2[\sum_{i=1}^K \int_0^{\infty}u_1w_iQ_i^{(1)}(du_1)]x_1+x_2^2-2[\int_0^{\infty}u_2w_aQ_a^{(2)}(du_2)-\sum_{i\neq a}\int_0^{\infty}u_2w_iQ_i^{(2)}(du_2)]x_2+const. \\
   & =x_1^2-2\sum_{i=1}^K v_i^{(1)}x_1+x_2^2-2[v_a^{(2)}-\sum_{i\neq a}v_i^{(2)}]x_2+const.
\end{align*}
where $v_i^{(j)}=\int_0^{\infty}uw_iQ_i^{(j)}(du)$.

To minimize $F(\textbf{x})$ is to minimize the quadratic form for $x_1,x_2$ and the solution is $x_1^*=\sum_{i=1}^3 v_i^{(1)}, x_2^*=v_a^{(2)}-\sum_{i\neq a}v_i^{(2)}$. Since $v_i^{(j)} \geq 0, x_1^*$ is always non-negative. Thus, one could apply C.L.T. on $x_1^*$ . However, for $x_2^*$, one have to discuss the three situations as did for the spider spaces. Therefore, in this case, if one separate the space into two directions, the stickiness of intrinsic mean would only occur on one direction, which is in a lower dimensional space. If for all a, $v_a^{(2)}<\sum_{i\neq a}v_i^{(2)}$, the intrinsic sample mean would stick to the spine S, but still has one degree of freedom. Thus for $n\to \infty,$ the intrinsic sample mean would stick to the spine S, and if it's coordinate on the spine is $\bar X_{I,n},$ than $\sqrt{n}(\bar X_{I,n},-\mu_I)\to_d Y,$ where $Y \sim \mathcal N(0,\sigma^2).$ Here $\mu_I$ is the mean of the projection of $Y$ on the spine, and $\sigma^2$ is its variance.

Another case of interest (see Barden et al(2013)\cite{BaLeOw:2013}), is data on a $Q_5$, as a union of five quadrants, that is isometric with a subset made of five quadrants on $T_4,$ whose trace on a the Petersen form a cycle ( see eg the cycle of quadrants from the edge $x$ to itself in Figure \ref{f:t4}. In this case the asymptotic distribution according to Barden et al (2013)\cite{BaLeOw:2013} (see also Barden and Le(2018)\cite{BaLe:2018}), behaves, via the piecewise Riemannian Log map, like a bivariate normal.

\section{Application to SARS-CoV-2}

\subsection{Background}

The SARS-CoV-2 was first identified in Wuhan, China, in 2019. As of June 13, 2021, at least 177 million  cases have been confirmed worldwide\cite{coviddata:2021}. As we speak, the number of cases keeps increasing. More than  220 countries and territories have been affected, resulting in at least 3,827,000 deaths; more than 161 million people have recovered.

The virus primarily spreads via respiratory droplets produced during breathing, coughing, sneezing, singing or talking. It may also be transmitted via contaminated surfaces. The time between exposure and symptom onset is typically five days, but may range from two to fourteen days. Symptoms may include fever, cough, and shortness of breath. By mid-December 2020, National regulatory authorities have approved six vaccines for public use: two  RNA based vaccines (tozinameran from Pfizer\&BioNTech and mRNA-1273 from Moderna), three conventional inactivated vaccines (Jenssen from Johnson $\&$ Johnson, BBIBP-CorV from Sinopharm and CoronaVac from Sinovac), and two viral vector vaccines (Gam-COVID-Vac from the Gamaleya Research Institute and AZD1222 from the University of Oxford and AstraZeneca). Recommended preventive measures include hand washing, maintaining distance from people, given tat there are asymptomatic cases, and monitoring and self-isolation for fourteen days if an individual suspects being infected.

\subsection{Data Processing}
\begin{figure}[H]
\centering
\includegraphics[width=0.8\linewidth]{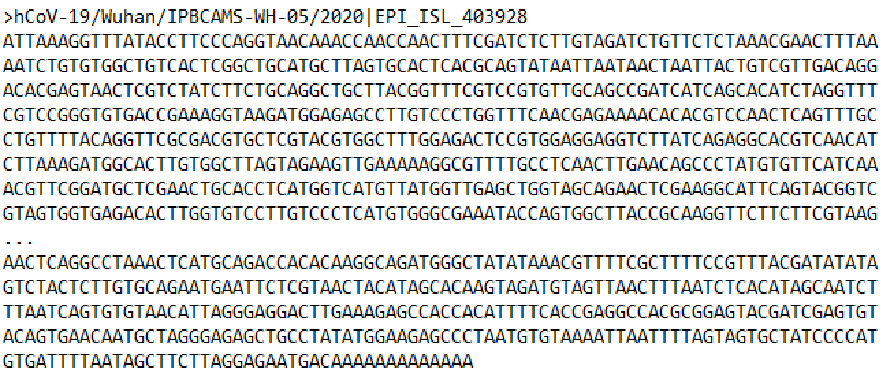}
\caption{Example of SARS-CoV-2 RNA subsequence data}%
\label{fig:rawdata}
\end{figure}

\begin{figure}[H]
\centering
\includegraphics[width=0.9\linewidth]{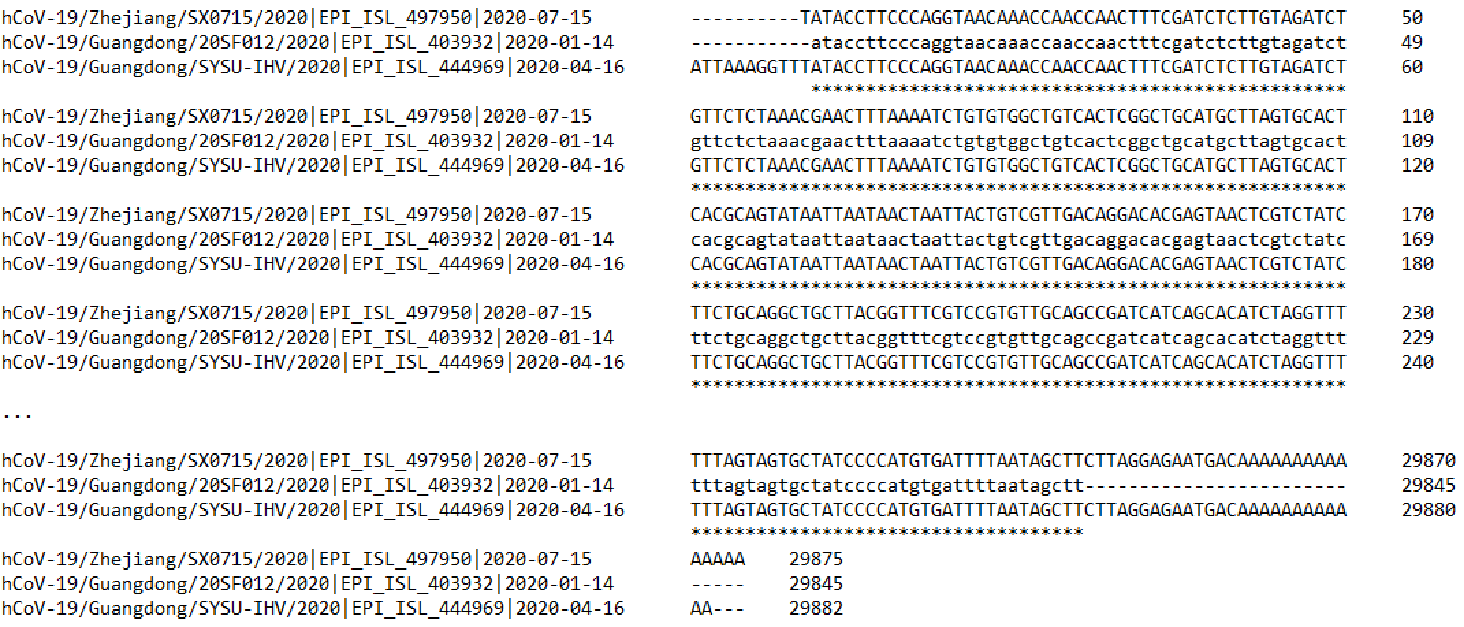}
\caption{Example of portion of aligned RNA sequences}%
\label{fig:aligned}
\end{figure}

The data we used in this paper are from multiple resources including GenBank\cite{HeZsByBoNeOySaBj:2017} for SARS-CoV\cite{SARS:2004}, MERS-CoV\cite{MERS:2012},and SARS-CoV-bat\cite{batSARS:2005} and GISAID\cite{EsBg:2017} for SARS-CoV-2\cite{SARS2:2020}. The data includes the virus name, host species, sample date and sample location. From the raw data green see figure \ref{fig:rawdata}, we first align the RNA sequences with each other via Clustal Omega method  \cite{SfWaDdGjKkLwLrMhRmSjTjHg:2011}. The Clustal Omega method uses the k-tuple distance to construct the distance matrix. The vector distances in the distance matrix are then used to cluster the sequences into subclusters using a k-means approach. From the aligned RNA sequences, see figure \ref{fig:aligned} , one could compute the fraction of mismatches at aligned positions, with gaps either ignored or counted as mismatches as distances from each RNA sequence. And then apply the {\em neighbor-joining method} \cite{SnNm:1987} to build trees.

\subsection{Comparison with other Coronaviruses}
In this part, we choose SARS-CoV, MERS-CoV and SARS-CoV-bat to compare with SARS-CoV-2.
SARS-CoV is the abbreviation for {\em severe acute respiratory syndrome coronavirus}, which  was an outbreak in June, 2003 with a global total of 8098 reported cases and 774 deaths, and a case fatality rate of $9.7\%$. The Middle East respiratory syndrome coronavirus (MERS-CoV) is another deadly coronavirus, which is currently not presenting a pandemic threat, and emerged in 2012, causing 2494 reported cases and 858 deaths in 27 countries and has a very high case fatality rate of $34\%$. SARS-CoV-bat is SARS-CoV from bats.

\begin{figure}[H]
\centering
\includegraphics[width=0.5\linewidth]{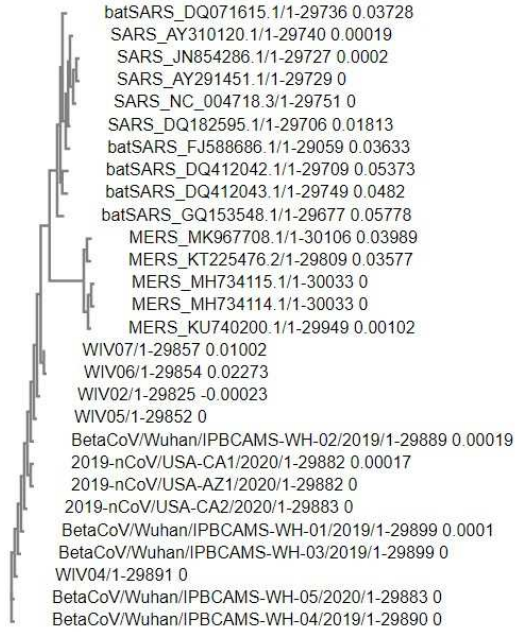}
\caption{Phylogenetic tree contains SARS-CoV, SARS-CoV-bat, MERS and SARS-CoV2 virus data}%
\label{fig:4virustree}
\end{figure}

We randomly pick 13 RNA sequences of SARS-CoV-2, 5 RNA sequences of SARS-CoV, 5 RNA sequences of SARS-CoV-bat and 5 RNA sequences of MERS-CoV. The tree containing all 28 samples could be built after alignment and distances computing, see figure \ref{fig:4virustree}.  From the figure \ref{fig:4virustree}, one could find that MERS viruses are clustered as one group; SARS and batSARS are mixed with each other; and SARS-2 viruses are generally different from MERS, SARS, and batSARS. To further analyze the relationship of the four viruses, we want to apply the tree spaces methods we introduced in Section 2, to build trees with four leaves and continue to compute mean tree to conclude a result.

To compare the gene information from each coronaviruses, we randomly pick one virus' RNA sequence from each group(SARS-CoV-2, SARS-CoV, MERS-CoV and SARS-CoV-bat) to build sample phylogenetic trees. The sample tree would be a tree with 4 leaves which contains SARS-CoV-2, SARS, MERS and batSARS.
\begin{figure}[H]
\centering
\includegraphics[width=0.6\linewidth]{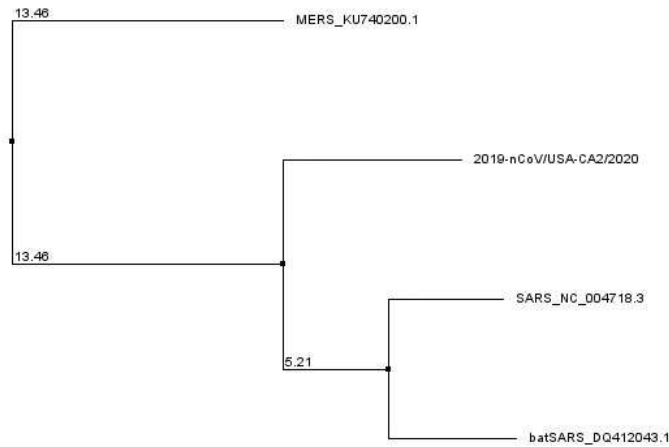}
\caption{The sample phylogenetic tree example}%
\label{fig:sampletree}
\end{figure}

Since all sample trees are in the same orthant (same book leaf of an open book), from CLT on tree spaces we discussed last section, we could apply multivariate normal distribution and compute the sample mean tree.

\begin{figure}[H]
\centering
\includegraphics[width=0.5\linewidth]{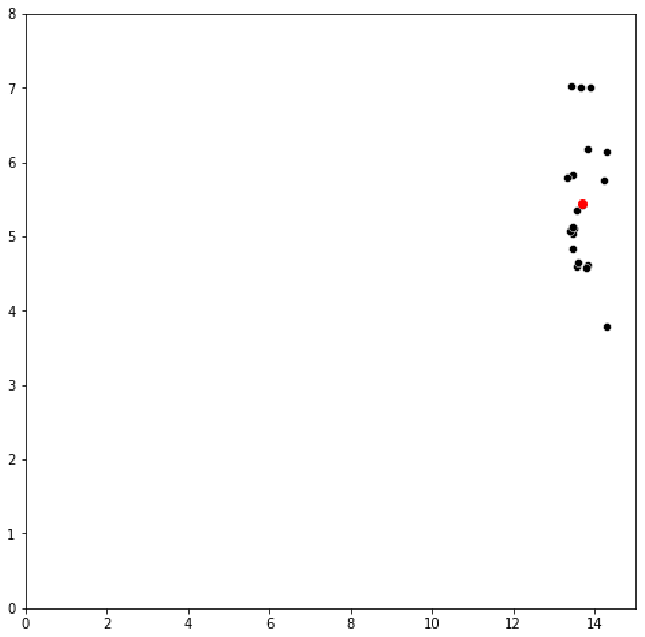}
\includegraphics[width=0.5\linewidth]{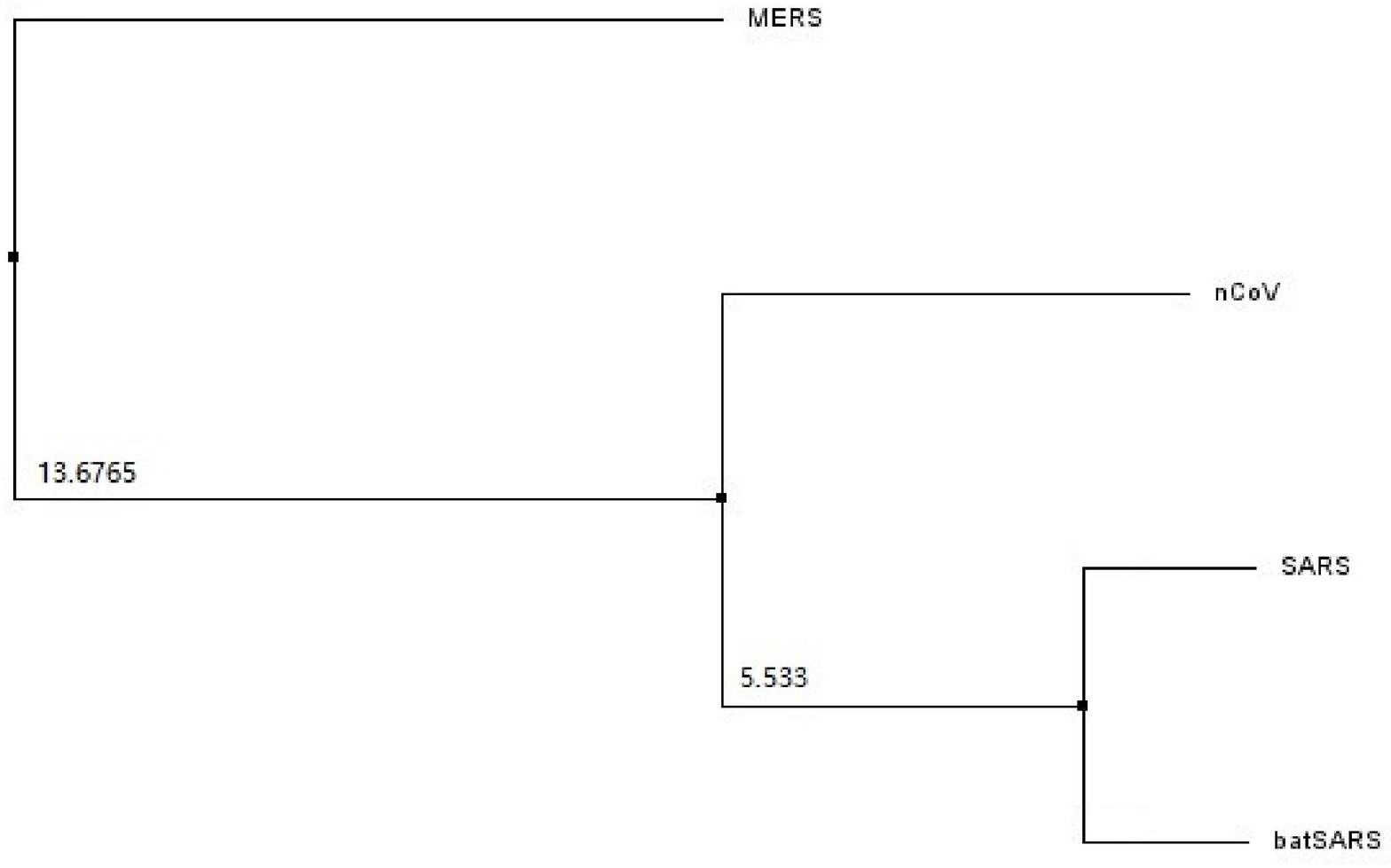}
\caption{Left:Sample trees on $T_4$ space, red dot: mean tree  Right: The phylogenetic mean tree}%
\label{fig:meantree}
\end{figure}

From the result, one could determine that the SARS-CoV-2's RNA sequencing is closer to SARS-CoV's and batSARS's. MERS's RNA sequencing is mostly different to SARS-CoV's and SARS-CoV-2's. The inneredge 13.67 and 5.53 represent the expected genetic distance from a virus to a possible common ancestor. eg. the inneredge 13.67 represents the expected genetic  distance between MERS and the common ancestor of SARS and SARS-2 is 13.67. Furthermore, based on the theorem introduced in the last section, one could compute a confidence region for the mean tree.

\subsection{Comparison with different months and locations on $T_3$ space}
As an RNA virus, the SARS-CoV-2 has high mutation rates, and these high rates are correlated with enhanced virulence and evolvability. Although researchers often assume that natural selection has optimized the mutation rate of RNA viruses from common knowledge, some data show that selection for faster replication is stronger and has a high possibility to make more mistakes. Moreover, mutation rates are evolvable and can respond to selection \cite{Duffy:2018}.

Since its emergence in late 2019, SARS-CoV-2 has diversified into several different co-circulating variants. Researchers have grouped the variants into different clades based on specific signature mutations. One of the most significant mutations is the D614G substitution in the gene coding for the ``spike" protein. This variant increases infectivity of SARS-CoV-2 in vitro and may have been selected by evolution or on purpose, for increased human-to-human transmission \cite{KORBER:2020812}.

Before countries closed borders at the early part of the pandemic, one could expect well-mixed SARS-CoV-2 phylogenetic tree structures across different countries and continents. Through March and April, many countries imposed differing types of 'lockdown' where the movement was restricted, and businesses and schools closed. Those restrictions decreased between-country transmission, making it more likely that different countries have different types of SARS-CoV-2 phylogenetic tree structures. However, with loose restrictions, and local and within-country transmission might make the tree structure more complicated.

Base on the information, we try to apply the phylogenetic tree space idea to analyze the mutations of SARS-CoV-2. We would apply whole RNA sequences to consider all beneficial, neutral, and deleterious mutations occurring in both the coding and non-coding regions. The data is obtained from GISAID\cite{EsBg:2017}, which contains 400 SARS-CoV-2 sequences from 10 countries. The data would be set to different groups based on the locations and months.

\begin{figure}[H]
\centering
\includegraphics[width=0.4\linewidth]{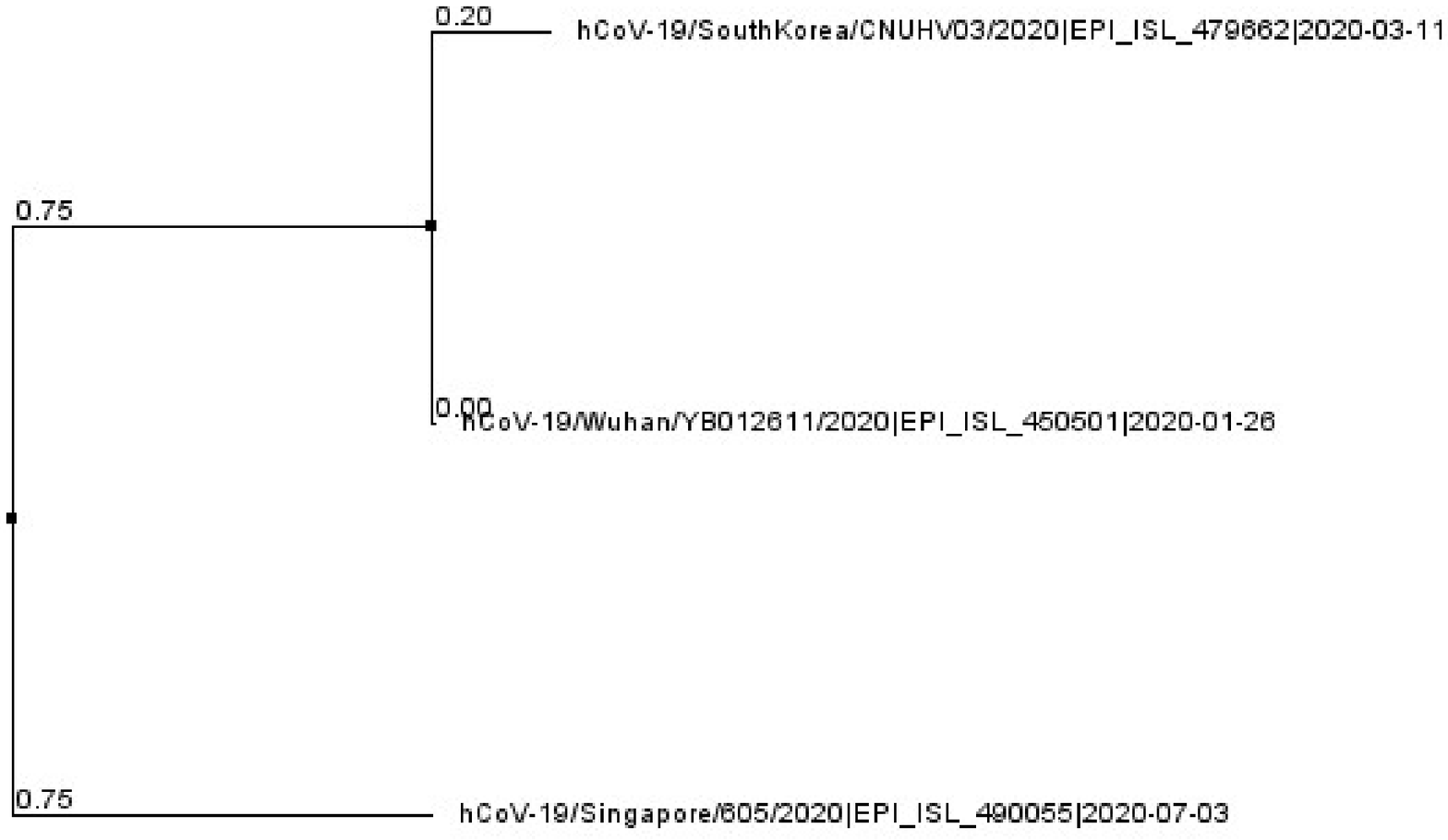}
\includegraphics[width=0.4\linewidth]{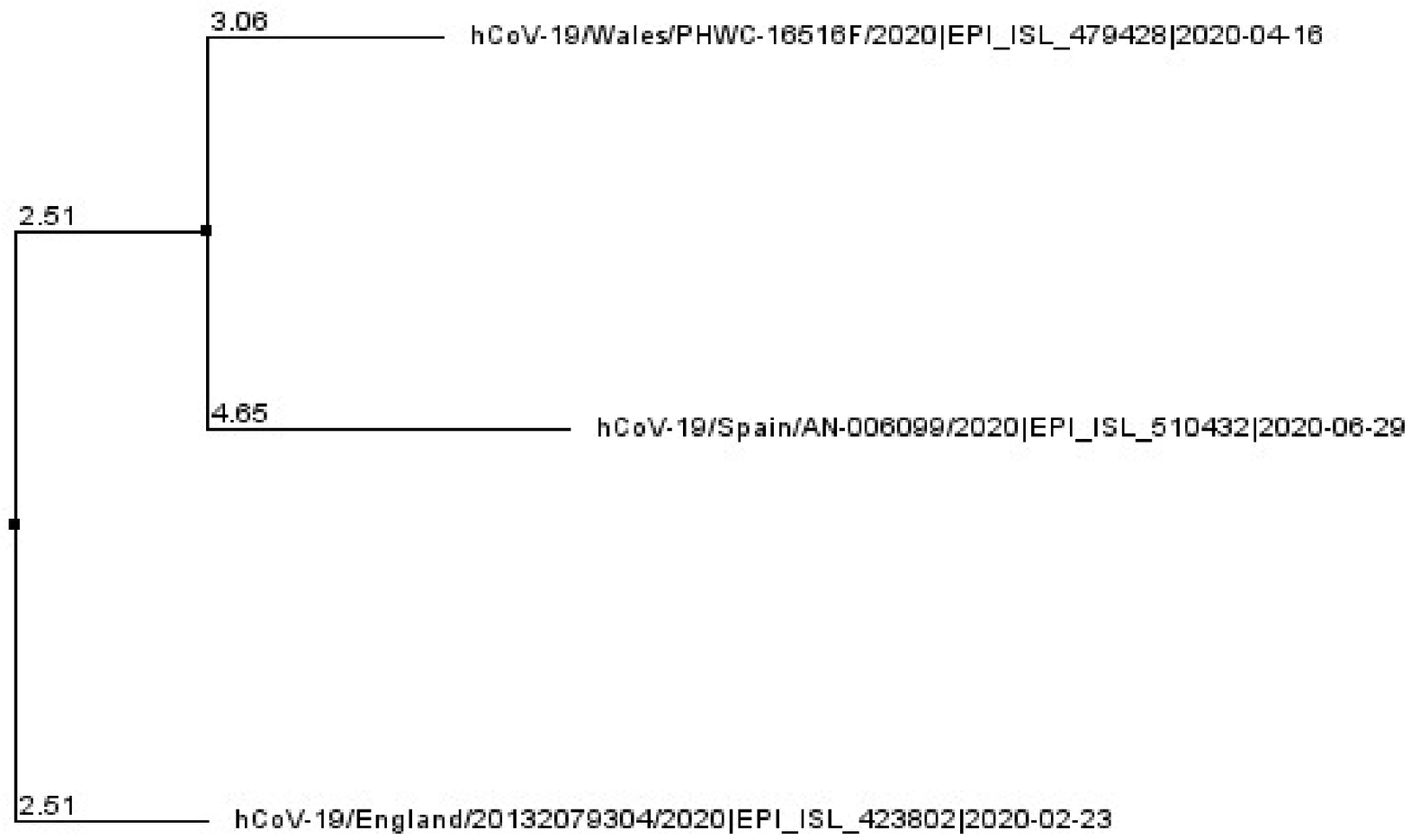}
\includegraphics[width=0.4\linewidth]{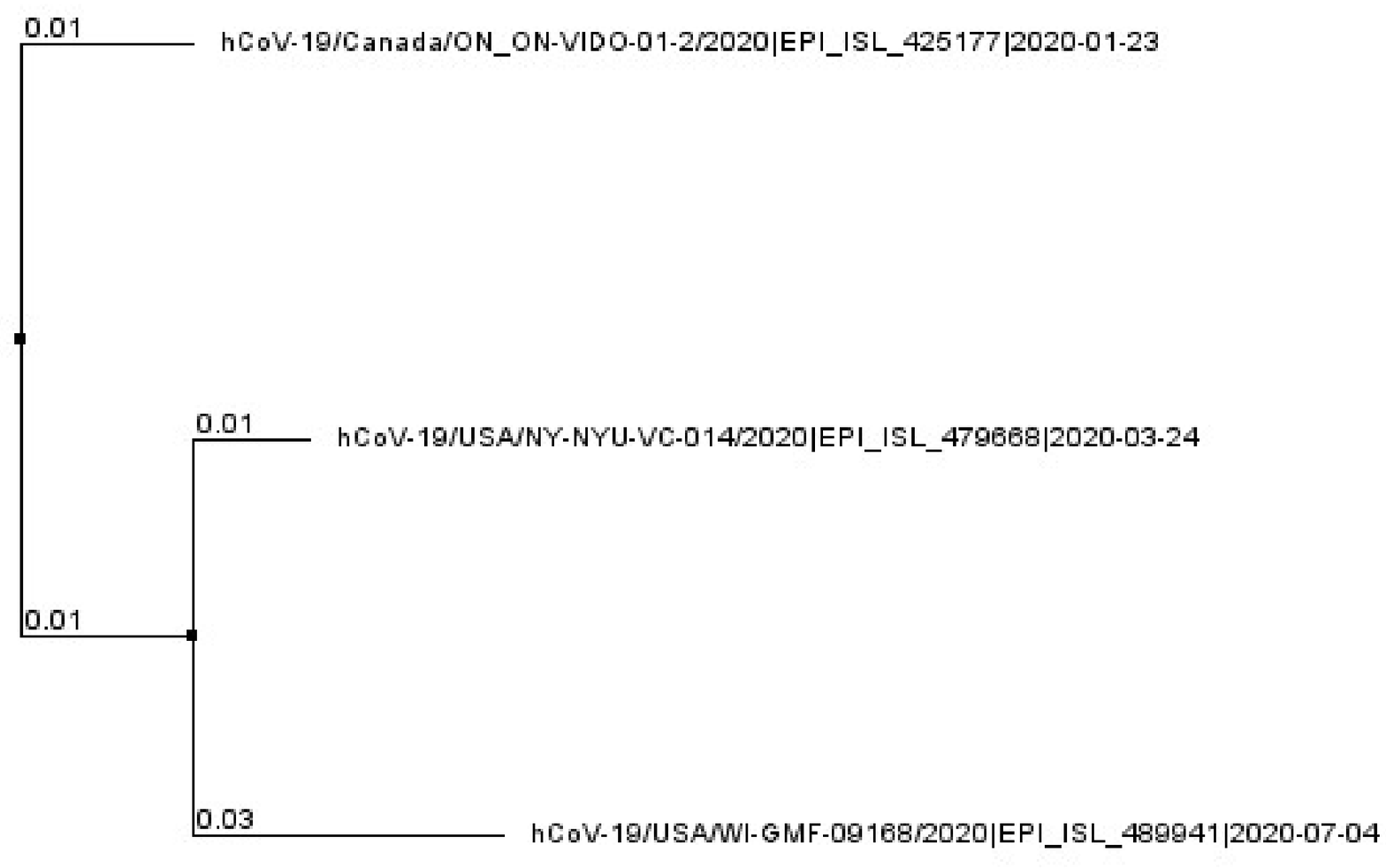}
\caption{The sample tree examples from left to right are from Asia, Europe and America}%
\label{fig:t3samples}
\end{figure}

First, we set 3 groups for each country. The groups are divided by different months: Jan-Mar, Apr-Jun, Jul-Sep. We randomly pick one RNA sequence from the three groups for each location and then make an alignment. We repeat the step 10 times to form aligned sequences samples for comparing viruses for different countries and 20 times for comparing viruses for different continents. From those aligned sequences, we could build trees with 3 leaves. One of the sample trees for each continent are as followed \ref{fig:t3samples},

\begin{figure}[H]
\centering
\includegraphics[width=0.6\linewidth]{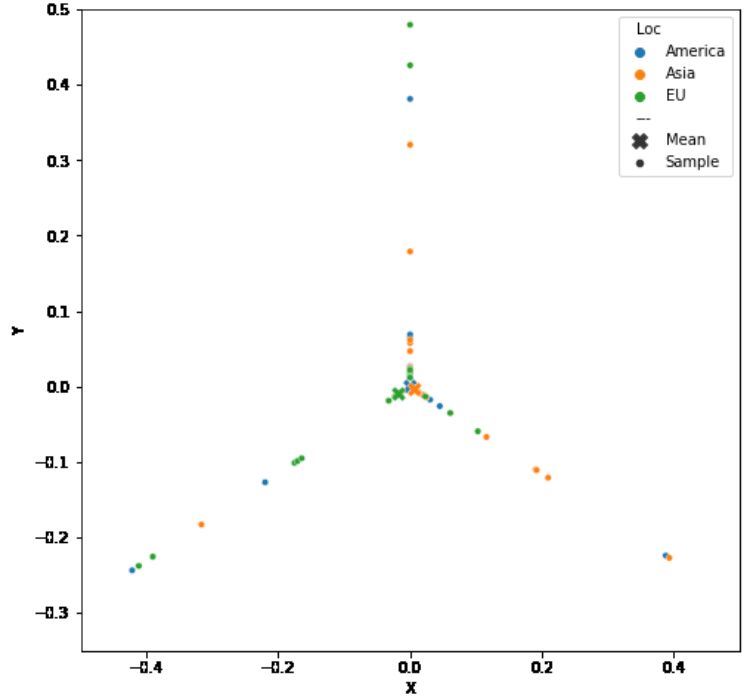}
\caption{Sample trees on $T_3$, Orange: Asia, Green: Europe, Blue: America }%
\label{fig:t3contient}
\end{figure}

Since our sample trees are on $T_3$ space, we could visualize all sample trees on a 3-spider, and we could compute the mean tree via methods we introduced in the last section. The figure \ref{fig:t3contient} showed the 20 sample trees for each continent and the corresponding mean trees.

The mean and standard deviation calculated are as followed. We use Newick format to present the tree type: 'a' represents the sample from group Jan-Mar, 'b': Apr-Jun, and 'c': Jul-Sep. From the figure, means are all very close to the origin or at the origin. The results show that the mean phylogenetic trees are close to each other and also close to the star tree, which means the difference between the viruses from different months is tiny. Another reason for mean trees closing to the origin might be the stickiness of the mean introduced in last section.

\begin{table}[h]
\centering
\begin{tabular}{l|lll}
        & Tree Type & Intrisic Mean  & Intrinsic Standard Deviation \\ \hline
America & (a,b,c)   & 0     & 0.259              \\
Asia    & ((a,c),b) & 0.008 & 0.256              \\
Europe  & ((b,c),a) & 0.02  & 0.302
\end{tabular}
\end{table}

We repeat the same steps for countries and have results showed as follows. From the figures below, one can find that mean trees for countries are not close to the origin except those from Asia. The mean tree stays on a particular 'leg' of the $T_3$ space could be expressed that there is a specific evolutionary or mutation pattern in that country. However, considering the standard deviations are not small enough to support a confident region locating on a particular 'leg' of the $T_3$ space. We still need more evidence to conclude significant results of a specific mutation pattern that occurred.

\begin{figure}[H]
\centering
\includegraphics[width=0.4\linewidth]{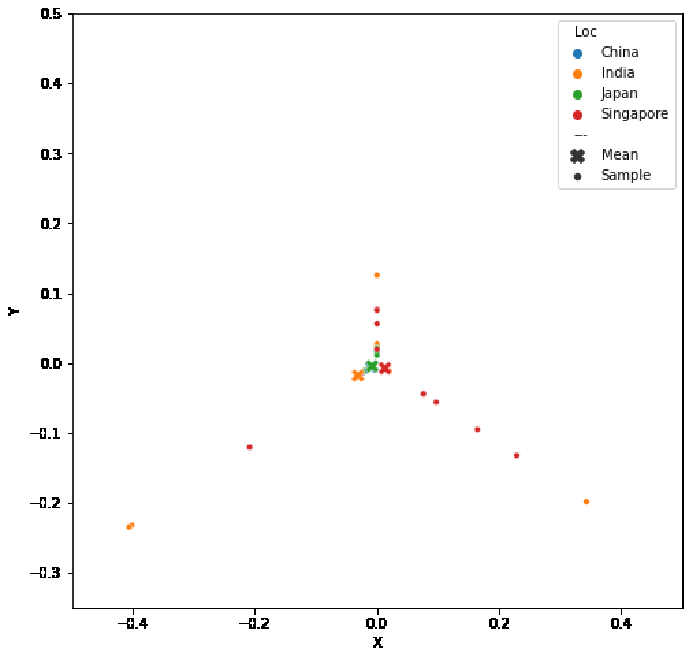}
\includegraphics[width=0.4\linewidth]{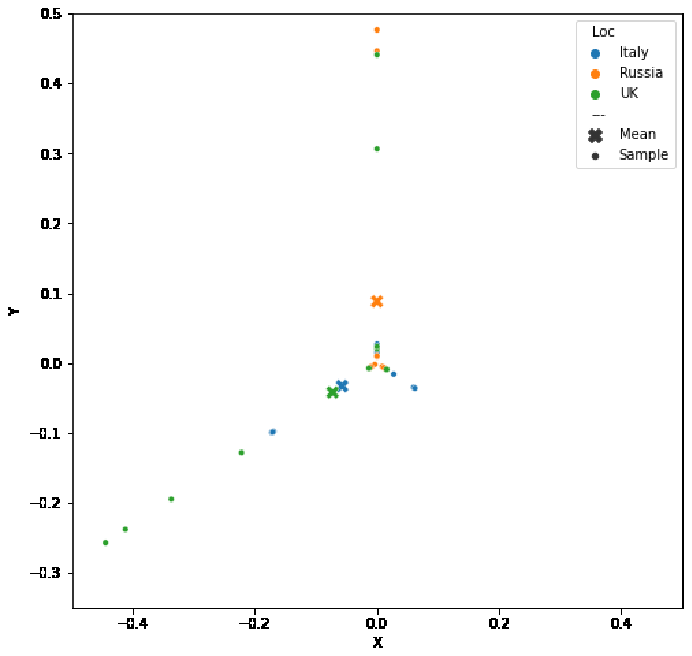}
\includegraphics[width=0.4\linewidth]{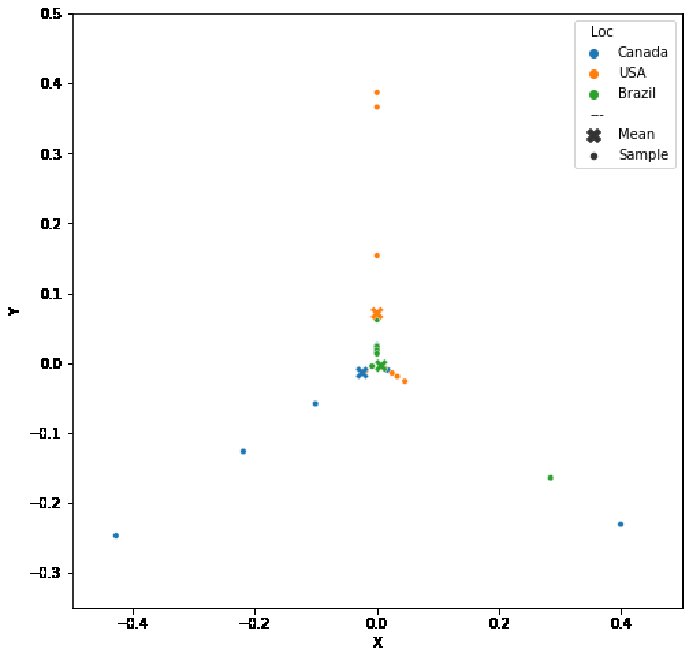}
\includegraphics[width=0.4\linewidth]{blank.eps}
\caption{Sample trees on $T_3$}%
\label{fig:t3countries}
\end{figure}

\begin{table}[H]
\centering
\begin{tabular}{llll}
\multicolumn{1}{l|}{}          & Tree Type & Intrisic Mean  & Intrinsic Standard Deviation  \\ \hline
\multicolumn{1}{l|}{China}     & ((b,c),a) & 0.01  & 0.022              \\
\multicolumn{1}{l|}{India}     & ((b,c),a) & 0.036 & 0.343              \\
\multicolumn{1}{l|}{Japan}     & ((b,c),a) & 0.01  & 0.018              \\
\multicolumn{1}{l|}{Singapore} & ((a,c),b) & 0.015 & 0.198              \\
\end{tabular}
\end{table}

\begin{table}[H]
\centering
\begin{tabular}{llll}
\multicolumn{1}{l|}{}          & Tree Type & Intrisic Mean  & Intrinsic Standard Deviation  \\ \hline
\multicolumn{1}{l|}{Italy}     & ((b,c),a) & 0.066 & 0.156              \\
\multicolumn{1}{l|}{Russia}    & ((a,b),c) & 0.088 & 0.274              \\
\multicolumn{1}{l|}{UK}        & ((b,c),a) & 0.084 & 0.44               \\
\end{tabular}
\end{table}

\begin{table}[H]
\centering
\begin{tabular}{llll}
\multicolumn{1}{l|}{}          & Tree Type & Intrisic Mean  & Intrinsic Standard Deviation  \\ \hline
\multicolumn{1}{l|}{Brazil}    & ((a,c),b) & 0.008 & 0.158              \\
\multicolumn{1}{l|}{Canada}    & ((b,c),a) & 0.028 & 0.322              \\
\multicolumn{1}{l|}{US}        & ((a,b),c) & 0.071 & 0.238
\end{tabular}
\end{table}

\subsection{Investigating the existence of a mean star 3-Spider for a 3-Spider plot}

For this application on the tree space $T_3$ , we consider CLT for the intrinsic mean on this tree space; according to Theorem \ref{t:clt-on-spider}, if the probability moment on one of the three spider legs of 3-spider does not exceed the sum of the probability moments on the other two spider legs, then the star tree (center of the spider) is the intrinsic mean . Biologically, this can imply the common ancestor, or indicate lack of significant DNA/RNA mutations. In particular, we will apply .
Therefore, our main aim here is to use the constructed $T_3$ tree space on the 3-spider to find out whether there has been significant RNA/DNA sequences mutations among the specified groups by making sure the overall sample mean for each tree is either non-sticky or sticky one. The overall mean inner edge values for all the three legs of the 3-spider satisfying  stickiness (star tree), implies no significant mutations among the sample groups being compared. So the 3-spider plot we got should either be a star spider or a non-star spider.

\subsection{Application of stickiness Theorem on SARS-CoV-2 RNA Sequence Data From N. America, Asia, and EU, With Their Specified Countries}

\begin{table}[hbt!]
\centering
\caption{Stickiness theorem on the three continents' 3-spider}
\begin{tabular}{|l|l|l|l|l|l|}
\hline
Legs & \multicolumn{3}{l|}{$w_a\nu_a - \sum_{b \neq a}w_b \nu_b=\theta$}                             & $\theta$ value & Verdict \\ \hline
1    & \begin{tabular}[c]{@{}l@{}}$\hat{w_a}=25/59$\\ $\hat{\nu_a}=2.3938$\end{tabular} & \begin{tabular}[c]{@{}l@{}}$\hat{w_{b_1}}=16/59$\\ $\hat{\nu_{b_1}}=2.134$\end{tabular}  & \begin{tabular}[c]{@{}l@{}}$\hat{w_{b_2}}=18/59$\\ $\hat{\nu_{b_2}}=2.8401$\end{tabular} & $-0.4$         &  \\ \hline
2    & \begin{tabular}[c]{@{}l@{}}$\hat{w_a}=16/59$\\ $\hat{\nu_a}=2.1342$\end{tabular} & \begin{tabular}[c]{@{}l@{}}$\hat{w_{b_1}}=25/59$\\ $\hat{\nu_{b_1}}=2.3938$\end{tabular} & \begin{tabular}[c]{@{}l@{}}$\hat{w_{b_2}}=18/59$\\ $\hat{\nu_{b_2}}=2.8401$\end{tabular} & $-1.3$         & Sticky  \\ \hline
3    & \begin{tabular}[c]{@{}l@{}}$\hat{w_a}=18/59$\\ $\hat{\nu_a}=2.8401$\end{tabular} & \begin{tabular}[c]{@{}l@{}}$\hat{w_{b_1}}=25/59$\\ $\hat{w_{b_1}}=2.3938$\end{tabular}   & \begin{tabular}[c]{@{}l@{}}$\hat{w_{b_2}}=16/59$\\ $\hat{\nu_{b_2}}=2.1342$\end{tabular} & $-0.7$         &   \\ \hline
\end{tabular}
\label{tab:sti1}
\end{table}

According to what the table \ref{tab:sti1} details in accordance with stickiness theorem \ref{t:clt-on-spider}, all three legs of the 3-spider satisfied the stickiness condition. Examination of the 3-spider plot in figure \ref{fig:t3contient} also seems to confirm this: All the means are clustered close to the center. Thus, the overall intrinsic sample mean for the continental regions of N.America, Asia and EU is the star tree. Therefore, by the third part of Theorem \ref{t:clt-on-spider}, we can conclude that SARS-CoV-2 virus RNA sequences from these regions did not mutate significantly, and thus, are not significantly different.
\begin{table}[hbt!]
\centering
\caption{Stickiness theorem on N. American countries' 3-spider}
\begin{tabular}{|l|l|l|l|c|l|}
\hline
Legs & \multicolumn{3}{l|}{$w_a\nu_a - \sum_{b \neq a}w_b \nu_b=\theta$}                                    & \multicolumn{1}{l|}{$\theta$ value} & Verdict    \\ \hline
1    & \begin{tabular}[c]{@{}l@{}}$\hat{w_a}=16/30$\\ $\hat{\nu_a}=1.2474$\end{tabular} & \begin{tabular}[c]{@{}l@{}}$\hat{w_{b_1}}=7/30$\\ $\hat{\nu_{b_1}}=0.9424$\end{tabular}  & \begin{tabular}[c]{@{}l@{}}$\hat{w_{b_2}}=7/30$\\ $\hat{\nu_{b_2}}=0.9395$\end{tabular} & $0.2$                               &  \\ \hline
2    & \begin{tabular}[c]{@{}l@{}}$\hat{w_a}=7/30$\\ $\hat{\nu_a}=0.9424$\end{tabular}  & \begin{tabular}[c]{@{}l@{}}$\hat{w_{b_1}}=16/30$\\ $\hat{\nu_{b_1}}=1.2474$\end{tabular} & \begin{tabular}[c]{@{}l@{}}$\hat{w_{b_2}}=7/30$\\ $\hat{\nu_{b_2}}=0.9395$\end{tabular} & $-0.7$                              & Not Sticky    \\ \hline
3    & \begin{tabular}[c]{@{}l@{}}$\hat{w_a}=7/30$\\ $\hat{\nu_a}=0.9395$\end{tabular}  & \begin{tabular}[c]{@{}l@{}}$\hat{w_{b_1}}=16/30$\\ $\hat{w_{b_1}}=1.2474$\end{tabular}   & \begin{tabular}[c]{@{}l@{}}$\hat{w_{b_2}}=7/30$\\ $\hat{\nu_{b_2}}=0.9424$\end{tabular} & $-0.7$                              &      \\ \hline
\end{tabular}
\label{tab:sti2}%
\end{table}

The figure \ref{fig:t3countries} (bottom) of the 3-spider of the N. American countries suggest there could be some mutation among their SARS-CoV-2 viruses. Going by the theorem \ref{t:clt-on-spider}, the result detailed in the table \ref{tab:sti2} revealed accordingly that the overall intrinsic mean is non-sticky and there is a high chance it is situated somewhere on leg 1 of the 3-spider. The sample size of 30 is fairly large to say $\sqrt{30}(\bar X_{30,I}-\mu_{I})$ is approximately normally distributed. Obviously, the intrinsic sample mean is not the star tree in this case. So, the SARS-CoV-2 RNA sequences from the countries of Brazil, Canada, and the USA did have some significant mutations among them. Therefore, we can declare, based on the theorem \ref{t:clt-on-spider}, that the SARS-CoV-2 RNA sequences from these N. American countries have some significant difference from each other. Further analysis may throw more light on the scope of this difference.

\begin{table}[hbt!]
\centering
\caption{Stickiness theorem on Asian countries' 3-spider}
\begin{tabular}{|l|l|l|l|c|l|}
\hline
Legs & \multicolumn{3}{l|}{$w_a\nu_a - \sum_{b \neq a}w_b \nu_b=\theta$} & \multicolumn{1}{l|}{$\theta$ value} & Verdict    \\ \hline
1    & \begin{tabular}[c]{@{}l@{}}$\hat{w_a}=12/30$\\ $\hat{\nu_a}=0.4853$\end{tabular} & \begin{tabular}[c]{@{}l@{}}$\hat{w_{b_1}}=7/30$\\ $\hat{\nu_{b_1}}=1.0976$\end{tabular}  & \begin{tabular}[c]{@{}l@{}}$\hat{w_{b_2}}=21/30$\\ $\hat{\nu_{b_2}}=1.5386$\end{tabular} & $-1.1$                              &      \\ \hline
2    & \begin{tabular}[c]{@{}l@{}}$\hat{w_a}=7/30$\\ $\hat{\nu_a}=1.0976$\end{tabular}  & \begin{tabular}[c]{@{}l@{}}$\hat{w_{b_1}}=12/30$\\ $\hat{\nu_{b_1}}=0.4853$\end{tabular} & \begin{tabular}[c]{@{}l@{}}$\hat{w_{b_2}}=21/30$\\ $\hat{\nu_{b_2}}=1.5386$\end{tabular} & $-1.0$                              &  Not Sticky   \\ \hline
3    & \begin{tabular}[c]{@{}l@{}}$\hat{w_a}=21/30$\\ $\hat{\nu_a}=1.5386$\end{tabular} & \begin{tabular}[c]{@{}l@{}}$\hat{w_{b_1}}=7/30$\\ $\hat{w_{b_1}}=1.0976$\end{tabular}    & \begin{tabular}[c]{@{}l@{}}$\hat{w_{b_2}}=12/30$\\ $\hat{\nu_{b_2}}=0.4853$\end{tabular} & $0.6$                               &  \\ \hline
\end{tabular}
\label{tab:sti3}%
\end{table}

The 3-spider plot found in figure \ref{fig:t3countries} (top left) suggest there might not be much mutation among the SARS-CoV-2 RNA sequences of the Asian countries of China, India, Japan, and Singapore.
However, applying theorem \ref{t:clt-on-spider}, we clearly see a contradiction to this observation as the Table \ref{tab:sti3} results clearly show. The ``net" verdict is that the overall intrinsic mean is not sticky and may possibly fall on leg 3. With a sample size of 30, $\sqrt{30}(\bar X_{30,I}-\mu_{I})$ may be approximated by a normally distribution with mean 0. Again, the intrinsic mean on the 3-spider here is unlikely to be the star tree. Thus, we can conclude by the theorem \ref{t:clt-on-spider} that the SARS-CoV-2 RNA sequences from these four Asian countries have some significant mutations among them. Further analysis is needed to investigate the exact location and magnitude of these mutations.

\begin{table}[hbt!]
\centering
\caption{Stickiness theorem on EU countries' 3-spider}
\begin{tabular}{|l|l|l|l|c|l|}
\hline
Legs & \multicolumn{3}{l|}{$w_a\nu_a - \sum_{b \neq a}w_b \nu_b=\theta$}                                & \multicolumn{1}{l|}{$\theta$ value} & Verdict    \\ \hline
1    & \begin{tabular}[c]{@{}l@{}}$\hat{w_a}=10/30$\\ $\hat{\nu_a}=1.7743$\end{tabular} & \begin{tabular}[c]{@{}l@{}}$\hat{w_{b_1}}=6/30$\\ $\hat{\nu_{b_1}}=0.2151$\end{tabular}  & \begin{tabular}[c]{@{}l@{}}$\hat{w_{b_2}}=14/30$\\ $\hat{\nu_{b_2}}=2.5628$\end{tabular} & $-0.6$                              &     \\ \hline
2    & \begin{tabular}[c]{@{}l@{}}$\hat{w_a}=6/30$\\ $\hat{\nu_a}=0.2151$\end{tabular}  & \begin{tabular}[c]{@{}l@{}}$\hat{w_{b_1}}=10/30$\\ $\hat{\nu_{b_1}}=1.7743$\end{tabular} & \begin{tabular}[c]{@{}l@{}}$\hat{w_{b_2}}=14/30$\\ $\hat{\nu_{b_2}}=2.5628$\end{tabular} & $-1.7$                              & Not sticky     \\ \hline
3    & \begin{tabular}[c]{@{}l@{}}$\hat{w_a}=14/30$\\ $\hat{\nu_a}=2.5628$\end{tabular} & \begin{tabular}[c]{@{}l@{}}$\hat{w_{b_1}}=10/30$\\ $\hat{w_{b_1}}=1.7743$\end{tabular}   & \begin{tabular}[c]{@{}l@{}}$\hat{w_{b_2}}=6/30$\\ $\hat{\nu_{b_2}}=0.2151$\end{tabular}  & $0.6$                               & \\ \hline
\end{tabular}
\label{tab:sti4}%
\end{table}

For the EU countries, it is interesting that the 3-spider plot, see figure \ref{fig:t3countries} (top right), and the application of theorem \ref{t:clt-on-spider}, see table \ref{tab:sti4}, all agree that there is some significant difference in the mutations among the SARS-CoV-2 RNA sequences from Italy, Russia, and the UK. Therefore the intrinsic mean tree here is not a star tree, and the overall intrinsic mean may likely fall on leg 3, with $\sqrt{30}(\bar X_{30,I}-\mu_{I})$ being approximately normally distributed. Again, further analysis is needed here to define the scope of this difference.

\subsection{Comparison with different months and locations on $T_4$ space}

Next, we want to separate the virus sequences into four groups to form phylogenetic trees with 4-leaves. We use the same data as the last section and make groups: Jan- Mar, Apr-May, Jun-Jul, and Aug-Sep. We randomly pick one RNA sequence from the four groups for each location and then make an alignment. We repeat the step 10 times to form aligned sequences samples for comparing viruses for different countries and 20 times for comparing viruses for different continents. From those aligned sequences, we could build trees with {four} leaves.

Since the sample trees are in $T_4$ space, we cannot visualize the whole $T_4$ space in $R^3$. However we could centrally project the points onto the Petersen graph to briefly visualize the samples. The figure below showed the 20 sample trees for each continent and the corresponding sample intrinsic mean trees.

\begin{figure}[H]
\centering
\includegraphics[width=0.8\linewidth]{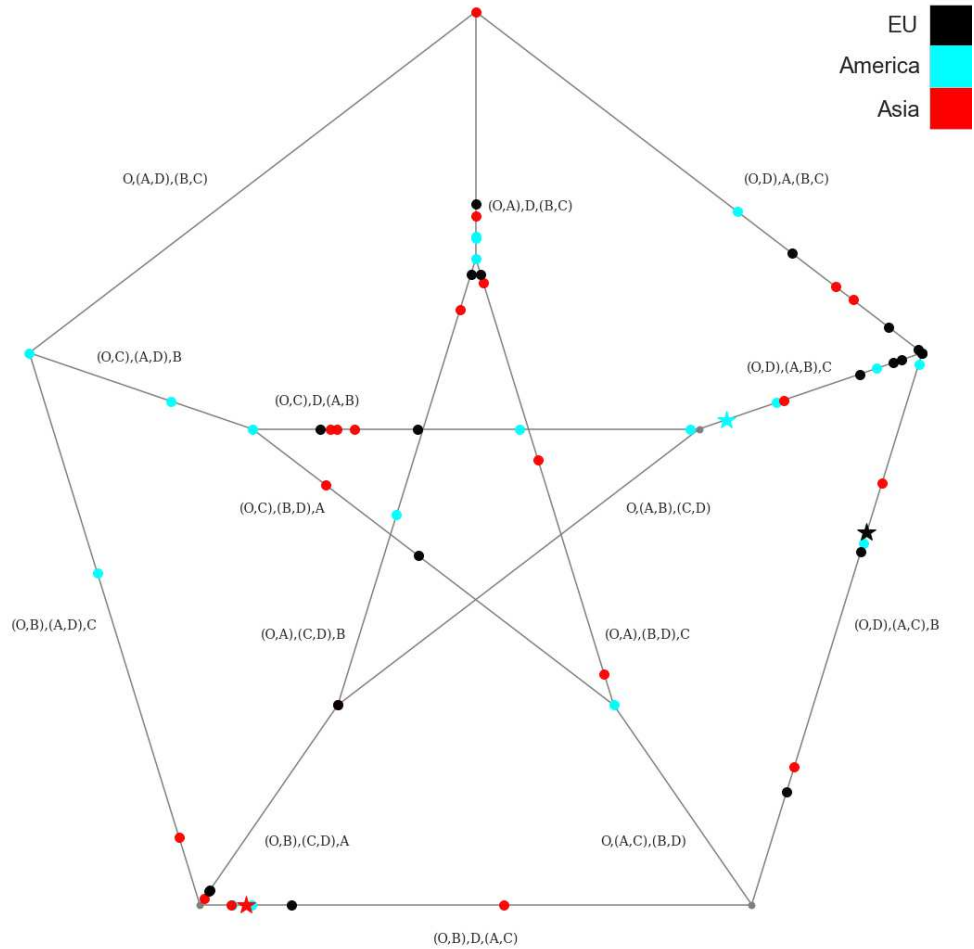}
\caption{Sample trees(dots) and mean trees(stars) of different continents on $T_4$ }%
\label{fig:t4contients}
\end{figure}

We could find that samples are separated in the $T_4$ space. The mean trees of the EU and America are close to each other rather than the mean tree of Asia. The mean trees of the EU and America show a similar mutation pattern, and there is a big difference with Asia's. However, we cannot draw any conclusion from the figure since the standard deviation and significant low value of inner edge. The mean tree's inner edge is low can be explained as the stickiness of the mean, and the sample size is small.

We repeat the same steps for countries and have results showed as follows,
\begin{figure}[H]
\centering
\includegraphics[width=0.8\linewidth]{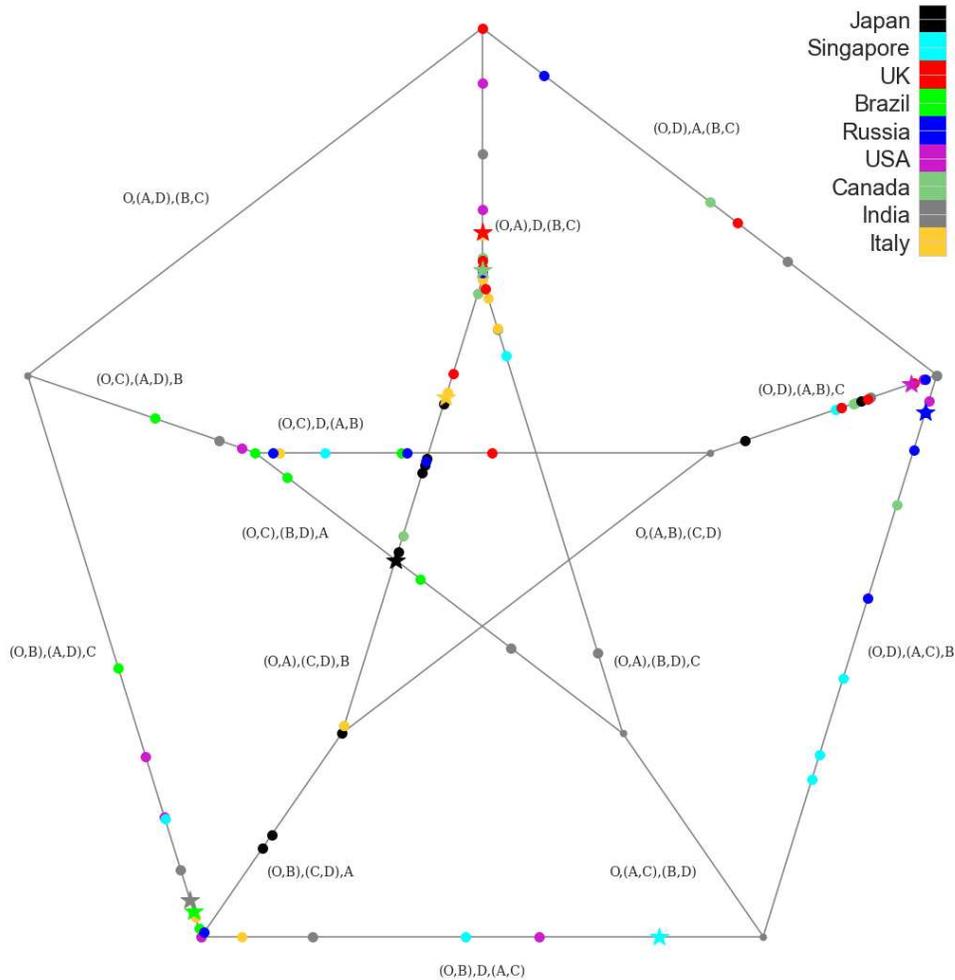}
\caption{Sample trees(dots) and mean trees(stars) of different countries on $T_4$}%
\label{fig:t4countries}
\end{figure}

The result shows that most sample trees locate on several particular orthants. Some countries share the same tree type and are very close to each other. However, as we mentioned before, we probably need more samples to draw a significant conclusion. As more data introducing, either more samples for each country or more countries added, we would find trends for the virus' evolutionary and mutation.


\end{document}